\documentclass[prd,reprint,superscriptaddress,preprintnumbers,nofootinbib,amsmath,amssymb,aps]{revtex4-2}

\usepackage{ads}
\usepackage{graphicx}
\usepackage{appendix}
\usepackage{dcolumn}
\usepackage[dvipsnames]{xcolor}
\usepackage[colorlinks=true,citecolor=blue,urlcolor=blue]{hyperref}
\usepackage{ulem}

\begin{document}

\title{Interacting light thermal-relic dark matter: self-consistent cosmological bounds}
\author{Rui An}
\thanks{Email: anrui@usc.edu}
\affiliation{Department of Physics and Astronomy, University of Southern California, Los Angeles, CA 90089, USA}
\author{Kimberly K.~Boddy}
\thanks{Email: kboddy@physics.utexas.edu}
\affiliation{Texas Center for Cosmology and Astroparticle Physics, Weinberg Institute, Department of Physics, The University of Texas at Austin, Austin, TX 78712, USA}
\author{Vera Gluscevic}
\thanks{Email: vera.gluscevic@usc.edu}
\affiliation{Department of Physics and Astronomy, University of Southern California, Los Angeles, CA 90089, USA}
\affiliation{TAPIR, Mailcode 350-17, California Institute of Technology, Pasadena, CA 91125, USA}

\begin{abstract}

We analyze cosmic microwave background (CMB) data to constrain the mass and interaction strengths of thermally-produced dark matter (DM) in a self-consistent manner, simultaneously taking into account the cosmological effects of its mass and interactions. The presence of a light thermal-relic particle contributes non-negligibly to the radiation density during Big Bang Nucleosynthesis (BBN), altering the light-element yields, as well as the the effective number of relativistic particle species. On the other hand, DM interactions with the Standard Model can affect distribution of matter in later universe. Both mass and interactions alter CMB anisotropy on sub-degree scales. To understand and quantify the interplay of these effects, we consider elastic DM-baryon scattering with a momentum-transfer cross section that scales as a power law of the relative velocity between the scattering particles. In the range of thermal–relic DM masses relevant for BBN ($\lesssim$ 20 MeV), we find that the reconstruction of the DM mass and the scattering cross section from the CMB data features strong degeneracies; modeling the two effects simultaneously increases the sensitivity of the CMB measurements to both fundamental properties of DM. Additionally, we study the effects of late-time residual annihilation of a light thermal relic and provide improved CMB constraints on the DM mass and annihilation cross section. To examine degeneracy between DM mass, cross section for elastic scattering with baryons, and annihilation cross section, we consider a specific case of DM with an electric and magnetic dipole moments. We present new, self-consistent cosmological bounds for this model and discuss implications for future searches.

\end{abstract}

\maketitle

\section{Introduction}

Dark matter (DM) accounts for a significant fraction of the matter in our Universe \cite{2009GReGr..41..207Z, 1970ApJ...159..379R, 2003ARA&A..41..645R, 2006ApJ...648L.109C, 2020A&A...641A...6P}, but its physical nature remains a mystery.
Standard cosmology features cold and collisionless DM (CDM), interacting only gravitationally with the Standard Model (SM) of particle physics; howerver, a range of beyond-CDM paradigms is explored in the literature and can lead to unique observational consequences \cite{1983PhLB..120..127P,1996PhR...267..195J,2009JCAP...07..004F,2009PhRvD..79k5016K,2013PDU.....2..139F,2014arXiv1402.5143H,2015PhRvL.115f1301D,2017arXiv170704591B}.

In the standard thermal freeze-out scenario, DM is in chemical equilibrium with the thermal bath at early times.
As the temperature drops below the DM mass, DM becomes nonrelativistic; its equilibrium number density drops exponentially, eventually rendering DM annihilation inefficient for maintaining chemical equilibrium \cite{1965JETP...21..656Z,1965PhL....17..164Z,1966PhRvL..17..712C,2005PhR...405..279B}.
The time of onset of exponential suppression in DM number density is primarily dictated by DM mass, while the freeze-out abundance of DM is governed by its annihilation rate to SM particles.
During this decoupling process, the contribution of DM to the entropy density of the Universe is transferred to the thermal bath, slowing down the cooling of the bath. Additionally, DM behavior as a radiation-like or matter-like fluid affects the expansion rate of the Universe.

For high DM masses, $\gtrsim 20$~MeV, the decoupling process occurs sufficiently early in cosmic history that there are no observable effects on the temperature evolution and expansion rate.
For DM masses between 10~keV and 20~MeV, the decoupling process occurs around the time of Big Bang Nucleosynthesis (BBN) and may affect standard BBN predictions through changes to the expansion rate, photon-to-baryon density ratio, and weak-interaction rates.
As a result, DM mass can affect the production of light chemical elements \cite{1986PhRvD..34.2197K, 2004PhRvD..70d3526S, 2004JPhG...30..279B} and the effective number of relativistic particle species \cite{2013PhRvD..87j3517S, 2014PhRvD..89h3508N, 2015PhRvD..91h3505N}.
For even lower DM masses $\lesssim 10$~keV, the decoupling process does not impact BBN, but the presence of relativistic DM still alters the expansion rate during BBN.
These effects that originate during the BBN era can be captured in the cosmic microwave background (CMB) anisotropies \cite{2004PhRvD..69b3509T,2008PhRvD..78d3509I,2006PhRvD..73f3528I,2004PhRvD..69h3002B,2015PhRvL.115i1301F,2013PhRvD..87h3008H} and provide some of the most stringent bounds on the mass of light thermal-relic DM \cite{2013JCAP...08..041B, 2014MmSAI..85..175S, 2014PhRvD..89h3508N, 2015PhRvD..91h3505N, 2019JCAP...02..007E, 2021arXiv210903246G,2022JCAP...07..002A,2020JCAP...01..004S,2022JCAP...07..002A}.

Apart from its mass, the non-gravitational interactions of DM with baryons can also affect CMB anisotropies in a more direct manner, leading to changes in the CMB power spectra.
For example, DM-baryon elastic scattering suppresses the clustering of matter in the Universe through DM-baryon momentum transfer, which is absent in the standard $\Lambda$CDM model \cite{2001PhLB..518....8B,2002PhRvD..66h3505B,2002dmap.conf..333B,2005A&A...438..419B,2014PhRvD..89b3519D,2018PhRvL.121h1301G,2018PhRvD..98h3510B,2019ApJ...878L..32N,2021ApJ...907L..46M,2021PhRvD.104j3521N, 2018PhRvD..97j3530X,2018PhRvD..98b3013S,2018PhRvD..98l3506B,2021PhRvD.104j3521N}.
Additionally, late-time (post freeze-out) residual annihilation of DM into SM particles \cite{2004PhRvD..70d3502C, 2005PhRvD..72b3508P, 2020A&A...641A...6P} injects energy into the plasma, potentially altering the recombination history and increasing the optical depth of CMB photons.%
\footnote{Late-time annihilation occurs if there is an abundance of both DM and anti-DM particles. This scenario is not applicable to asymmetric DM \cite{2009PhRvD..79k5016K}.}
In summary, the fundamental properties of DM affect cosmological observables in three distinct ways: the mass of DM controls the onset of its decoupling from the thermal bath in the early Universe; the DM-baryon scattering cross section quantifies the rate of momentum transfer and affects clustering of matter; and the DM annihilation cross section determines the energy injection into the plasma at late times, as well as the DM relic abundance at the time of freeze-out.

The effects of DM mass and interactions are typically considered separately in cosmological data analyses.
However, under the assumption of a thermal-relic scenario, these effects can be simultaneously relevant to the same observables and their joint consideration may alter cosmological bounds on individual parameters.
For example, DM-baryon scattering at the level of current cosmological bounds typically implies that DM is in equilibrium with the thermal bath during BBN \cite{2014PhRvD..89b3519D}; thus, the effects of DM mass and scattering with baryons should be considered and analyzed together.
When considering late-time residual DM annihilation, the effects of DM mass should be included in the analysis for a consistent interpretation of limits on thermal-relic DM.
Additionally, in the simplest scenarios, DM scattering and annihilation originate from a single interaction term in the Lagrangian and the relevant cross sections are related to each other due to a crossing symmetry; however, if different interactions dominate the scattering and annihilation rates, the corresponding cross sections are independent parameters.

In this work, we present the first self-consistent cosmological likelihood analyses of the effects of DM mass and its interactions using CMB measurements from \textit{Planck} 2018 \cite{2020A&A...641A...6P}.
We derive the first joint bounds on the i) DM mass and DM-baryon scattering cross section, ii) DM mass and late-time annihilation cross section, and iii) DM mass, DM-baryon scattering cross section, and late-time annihilation cross section for the specific scenario of DM that possesses magnetic and electric dipole moments.
Case iii expands upon previous work, which did not account for the effects of DM mass and did not consider the joint effects of the electric and magnetic dipole interactions \cite{2004PhRvD..70h3501S,2006PhRvD..73h9903S,2021JHEP...11..156H}.
In all cases, we assume that DM achieves its relic abundance by annihilating into SM particles in the early Universe, but we do not enforce that this annihilation at freeze-out is tied to the elastic scattering or late-time annihilation cross section.
We find that in the range of thermal-relic masses relevant for BBN ($\lesssim 20$~MeV), the effects of mass and DM interactions can be degenerate and modeling them simultaneously often leads to an improvement in the sensitivity of CMB measurements to DM parameters.

This paper is organized as follows.
In Section~\ref{sec:theory}, we review how light thermal-relic DM alters standard BBN predictions, which in turn affect CMB anisotropies, and how DM-baryon scattering and residual DM annihilation alter CMB power spectra.
We then discuss the impact of the combined effects on CMB in Section~\ref{subsec:jointeffects}.
In Section~\ref{sec:constraints}, we describe the details of our methods and analyses, and we present the resulting constraints from \textit{Planck} data.
Finally, we summarize in Section~\ref{sec:summary}. 

\section{Thermal-relic DM}\label{sec:theory}

In this section, we review the impact of light thermal-relic DM on standard BBN predictions, which in turn affect CMB anisotropies, as well as the effects of DM-baryon scattering and late-time residual DM annihilation.

We assume that the observed DM abundance is achieved through a thermal freeze-out mechanism \cite{2017tasi.conf..399L,2019arXiv190407915L}, and determined by the specific details of the freeze-out process.
In this scenario, DM maintains thermal equilibrium through its interactions with SM particles in the early Universe, until the temperature of the thermal bath drops below the mass of DM.
However, in our analyses we do not require that the freeze-out processes that set the relic abundance are the same as those that are active at later times---deviations from the standard Lee-Weinberg scenario~\cite{Lee:1977ua} arise, e.g., in models of coannihilation~\cite{Griest:1990kh}, coscattering~\cite{Garny:2017rxs,DAgnolo:2017dbv}, and semi-annihilation~\cite{DEramo:2010keq}.

\subsection{Impact on BBN}\label{sec:BBN}

The abundances of light elements are established during BBN in the very early Universe, at temperatures in the range of $10\ \mathrm{keV} \lesssim T \lesssim 10\ \mathrm{MeV}$.
Since DM freeze-out occurs at temperatures near the DM mass, sufficiently heavy DM is nonrelativistic during BBN and thus contributes negligibly to the overall energy density of the Universe at that time.
In this case, DM does not interfere with standard BBN predictions.
However, for DM masses $m_\chi \lesssim 20$~MeV, DM is either relativistic throughout BBN or becomes nonrelativistic and freezes out during BBN, both of which impact BBN processes.

One of the main effects of the presence of light thermal-relic DM is to change the expansion rate in the early Universe, modifying the time at which proton--to--neutron conversion freezes out and therefore changing the values of primordial element abundances, compared to standard BBN predictions.
The resulting DM annihilation into SM particles can additionally affect the radiation content in the Universe, altering the effective number of light, neutrino-like species $N_\mathrm{eff}$, given by
\begin{equation}
  N_\mathrm{eff}(m_{\chi}) \equiv 3.044\left[\frac{11}{4}\left(\frac{T_\nu}{T_\gamma}\right)_0^3\right]^\frac{4}{3}
\end{equation}
which includes the contribution from the SM neutrinos \cite{2020JCAP...12..015F,2021JCAP...04..073B}, as well as contribution from light DM.
$(T_{\nu}/T_{\gamma})_{0}$ is the present-day ratio of neutrino-to-photon temperature.
For example, DM annihilating into photons after neutrino decoupling can heat photons relative to the decoupled neutrinos, leading to a reduced value of $N_\mathrm{eff}$; conversely, DM annihilating to neutrinos heats up neutrinos relative to photons, increasing $N_\mathrm{eff}$.

In previous work \cite{2022JCAP...07..002A}, we discussed the effects of a light DM mass $m_{\chi}$ on primordial abundances and on $N_\mathrm{eff}$ in two specific scenarios: one in which DM couples electromagnetically to the SM and another in which it only couples to SM neutrinos.
In each case, we considered DM to be a real scalar, complex scalar, Majorana fermion, or Dirac fermion.
The impact of these four types of DM particle on BBN are similar to each other; thus, in this work, we assume DM is an electromagnetically coupled Majorana fermion, as an illustrative example.

The effects of light thermal-relic DM on $N_\mathrm{eff}$ and the primordial helium mass fraction $Y_\mathrm{p}$ can be captured by the CMB anisotropy \cite{2004PhRvD..69b3509T,2008PhRvD..78d3509I,2006PhRvD..73f3528I,2004PhRvD..69h3002B,2015PhRvL.115i1301F,2013PhRvD..87h3008H}.
A larger $N_\mathrm{eff}$ leads to a faster expansion rate, resulting an increased Silk damping and thus suppressing the CMB power spectrum at small scale.
Conversely, a smaller $N_\mathrm{eff}$ leads to an enhanced small-scale CMB anisotropy.
The effect of $Y_\mathrm{p}$ on the CMB comes mainly from its impact on the recombination history, which changes the structure of acoustic peaks via diffusion damping at small scales.
Since helium-4 recombines before hydrogen, a larger $Y_\mathrm{p}$ corresponds to fewer free electrons available for hydrogen recombination; thus, the mean free path of the photons is larger, resulting in a larger damping length scale and suppressed small-scale anisotropy in the CMB.
Conversely, the CMB power spectrum is less damped for smaller values of $Y_\mathrm{p}$.

Previous studies have placed lower limits on the light DM mass using the measurements of CMB from the \textit{Planck} satellite, as well as other ground-based observations \cite{2013JCAP...08..041B, 2014MmSAI..85..175S, 2014PhRvD..89h3508N, 2015PhRvD..91h3505N, 2019JCAP...02..007E, 2021arXiv210903246G,2022JCAP...07..002A}.
For example, Ref.~\cite{2022JCAP...07..002A} constrained an electromagnetically coupled Majorana fermion with mass $m_{\chi} \gtrsim 4.85$ MeV at the 95\% confidence level (CL) using \textit{Planck} data alone. In addition to DM mass, non-gravitational interactions between DM and baryons can also alter CMB anisotropies in a more direct manner, leading to changes in CMB power spectra.
Under the assumption of a thermal-relic case, the effects of DM mass and interactions can be simultaneously relevant to the CMB observables, which may alter the cosmological bounds on individual parameters.
In this work, we focus our analysis on two specific DM interaction scenarios: DM-baryon scattering and late-time residual DM annihilation, as discussed in the following.

\subsection{DM-baryon scattering}\label{sec:IDM}

\begin{figure}[t]
\includegraphics[width=0.44\textwidth]{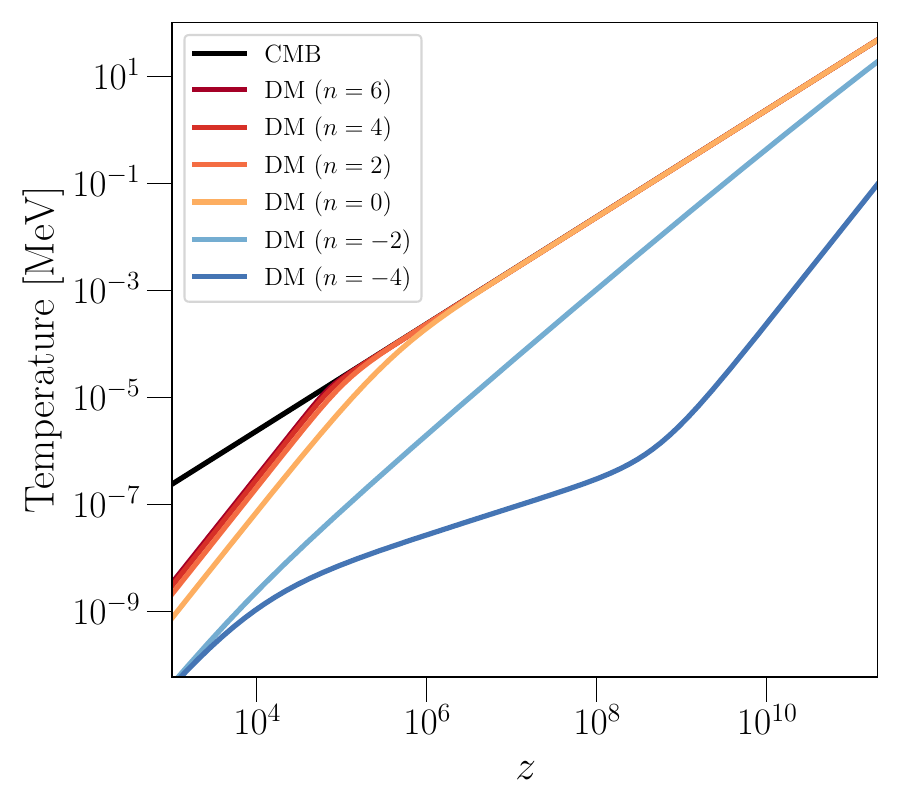}\\
\includegraphics[width=0.44\textwidth]{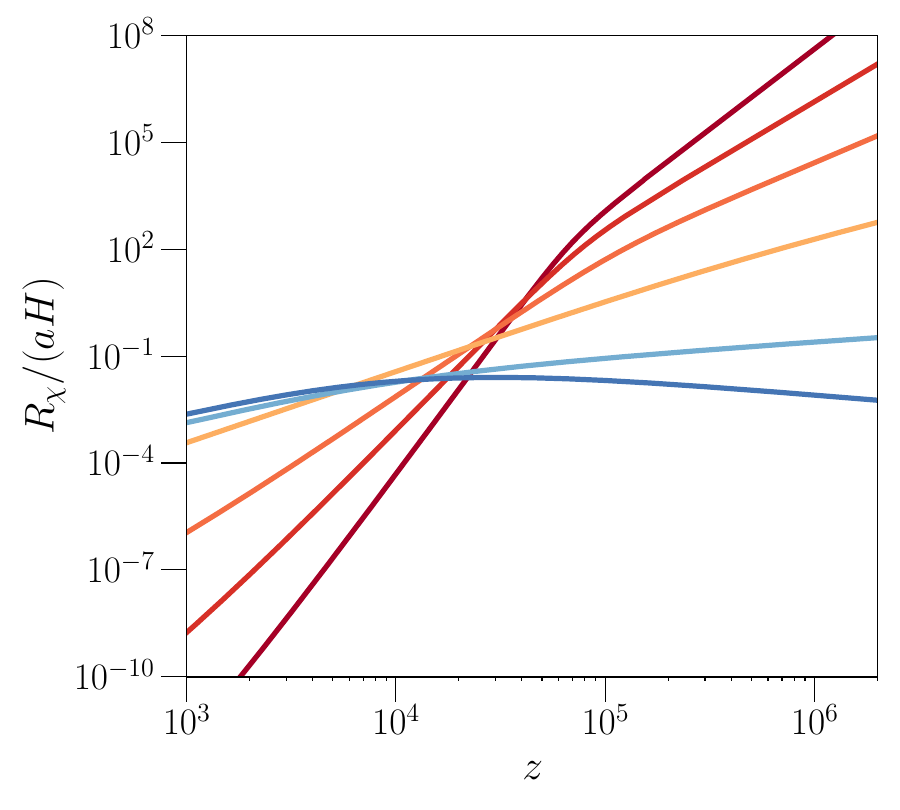}
\caption{Evolution of the DM temperature (\textit{upper panel}) and the ratio of the momentum-transfer rate coefficient $R_{\chi}$ to the expansion rate of the Universe (\textit{lower panel}) for different values of $n$ (colored lines). The DM mass is fixed at $m_{\chi} = 10$ MeV and the cross section $\sigma_0$ for each $n$ is set to its CMB 95\% CL upper bound from Ref.~\cite{2021PhRvD.104j3521N}. We also show the CMB photon temperature (black line) for reference in the upper panel; note that the evolution at very large $z$ is illustrative and assumes temperature scales simply as $1+z$. We properly account for changes in the temperature evolution from entropy conservation in our BBN analyses.}
\label{fig:Tdm}
\end{figure}

Elastic scattering between DM and baryons permits energy and momentum transfer between the DM and baryon fluids, which suppresses the formation of structure at progressively smaller scales, altering the shape of the CMB power spectra and of the matter power spectrum \cite{2001PhLB..518....8B,2002PhRvD..66h3505B,2002dmap.conf..333B,2005A&A...438..419B,2014PhRvD..89b3519D,2018PhRvL.121h1301G,2018PhRvD..98h3510B,2019ApJ...878L..32N,2021ApJ...907L..46M,2021PhRvD.104j3521N}.
Previous studies have placed upper limits on the momentum-transfer cross section as a function of DM mass ($\gtrsim 1$ keV) using the measurements of CMB anisotropies \cite{2014PhRvD..89b3519D,2018PhRvL.121h1301G,2018PhRvD..98h3510B,2018PhRvD..97j3530X,2018PhRvD..98b3013S,2018PhRvD..98l3506B,2021PhRvD.103d3541A,2015PhRvL.115g1304A,2015PhRvD..92h3528M,2022PhRvD.106d3510B} and a variety of other observational probes of structure \cite{2013PhRvD..88d3502V,2017PhRvD..96b3522I,2020MNRAS.492.3047H,2020MNRAS.491.6077G,2018JCAP...07..061B,2014MNRAS.442.2487K,2018MNRAS.473.2060J,2021PhRvL.126i1101N,2021JCAP...08..062N}.

We parameterize the momentum-transfer cross section between DM and baryons as $\sigma_\mathrm{MT} = \sigma_0 v^n$, where $\sigma_0$ is a constant coefficient and $v$ is the relative particle velocity with a power-law index $n$ \cite{2004PhRvD..70h3501S,2013PhRvD..88k7701D,2018PhRvD..98j3529K,2014PhRvD..89b3519D,2018PhRvD..98h3510B,2017PhRvD..96k5021K,2022PhRvD.106d3510B}.
This parameterization of the velocity dependence encompasses a wide class of DM models, and we consider several values of $n$ that are well-motivated: $n = \{-4, -2, 0, 2, 4, 6\}$.
The time evolution for the rate of momentum transfer $R_\chi$ depends on how the interaction cross section scales with the relative particle velocity. Here we concentrate on the studies of DM–proton scattering, with $R_\chi$ defined as~\cite{2014PhRvD..89b3519D}
\begin{equation}
    R_{\chi} = \frac{ac_n\rho_{\mathrm{p}}\sigma_{0}}{m_{\chi}+m_\mathrm{p}}\left(\frac{T_{\chi}}{m_{\chi}}+\frac{T_{\mathrm{b}}}{m_{\mathrm{p}}}+\frac{V^{2}_{\mathrm{RMS}}}{3}\right)^{-\frac{n+1}{2}},
\end{equation}
where $a$ is the scale factor, $c_n \equiv 2^{\frac{n+5}{2}}\Gamma(3+\frac{n}{2})/(3\sqrt{\pi})$, $\rho_p$ is the proton energy density, $m_\mathrm{p}$ is the proton mass, $T_{\chi}$ and $T_b$ denote DM and baryon fluid temperatures respectively, $V_{\mathrm{RMS}}^{2}$ represents the root-mean-square bulk relative velocity between DM and baryons, which is set to 30 for $n<0$ and 0 for $n\geq 0$.

The effects of scattering can be important at different epochs, sensitive to distinct cosmological probes \cite{2002astro.ph..2496C,2004PhRvD..70h3501S,2014PhRvD..89b3519D,2018PhRvL.121h1301G,2018PhRvD..98h3510B,2018PhRvD..97j3530X, 2018PhRvD..98l3506B,2014PhRvD..90h3522T,2015PhRvD..92h3528M,2018Natur.555...71B,2022PhRvD.106d3510B}. We show the evolution of the DM temperature and the momentum-transfer rate $R_\chi$, normalized by the expansion rate of the Universe, all way up to the BBN epoch ($z\sim 10^9$) in Fig.~\ref{fig:Tdm} for different choices of $n$. 
We fix $m_{\chi} = 10$ MeV and set $\sigma_0$ for each $n$ to its respective 95\% CL upper limit derived from a previous CMB analysis \cite{2021PhRvD.104j3521N}, keeping other standard cosmological parameters at their no-scattering best-fit \textit{Planck} 2018 values \cite{2020A&A...641A...6P}.
Larger values of $n$ lead to later thermal decoupling times of the DM fluid.
For models with $n \geq 0$, the scattering (with values of $\sigma_0$ near current CMB limits) is significant in the pre-recombination Universe \cite{2014PhRvD..89b3519D,2018PhRvD..98h3510B} and sufficient for DM to maintain thermal equilibrium with the thermal bath during BBN.
Therefore, we must also consider the constraints on DM mass that arise from light thermal-relic particles that impact BBN.

For models with $n \geq -2$, the rate of momentum transfer between DM and baryons is larger at higher redshifts \cite{2014PhRvD..89b3519D,2018PhRvD..98l3506B}.
However, for models with $n=-2$ and $n=-4$, assuming the benchmark value of 
$\sigma_0$ consistent with the current upper bounds (used in Fig.~\ref{fig:Tdm}), the interaction rate implied at the highest redshift of interest is not sufficient to achieve kinetic equilibrium. 
Therefore, constraints from BBN are not immediately applicable for $n=-2$ and $n=-4$, and we do not consider these individual cases in the first part of our analysis for this reason; however, in Section~\ref{sec:dipoleDM}, we do include $n=-2$ in the context of a specific model that thermalizes DM at early times through an $n=0$ interaction.

\subsection{Residual DM annihilation}\label{sec:Anni}

After DM freeze-out, there is still a small amount of residual annihilation that occurs through the same annihilation process that established relic abundance. However, in this work, we allow for the possibility that the annihilation process that sets the DM relic abundance during freeze-out differs from the annihilation process relevant at late times.
The residual annihilation injects energy into the thermal bath and can alter the ionization history of the Universe \cite{1991NuPhB.360..145G,2012PhRvD..86b3506S, 2004PhRvD..70d3502C, 2005PhRvD..72b3508P}.
During the epoch of recombination, DM annihilation produces high energy photons and electrons, which heat and ionize the hydrogen and helium gas as they cool. As a result, ionization fraction after recombination is larger, increasing the width of last scattering surface and consequently the width of the visibility function.
The broader last scattering surface damps correlations between temperature fluctuations and enhances low-$\ell$ correlations between polarization fluctuations.

The bound on late-time DM annihilation from \textit{Planck} 2018 data is $p_\mathrm{ann} < 3.5 \times 10^{-28}~\mathrm{cm^3/s/GeV}$ at 95\% CL \cite{2020A&A...641A...6P}, where
$p_\mathrm{ann} \equiv f_\mathrm{eff}\left\langle \sigma v \right\rangle /m_{\chi}$ is the effective parameter constrained by CMB anisotropies, where $\left\langle \sigma v \right\rangle$ is thermally averaged annihilation cross section, and $f_\mathrm{eff}$ is the fraction of the energy injected by the annihilation process that is transferred to the intergalactic medium around the redshifts that CMB anisotropies are most sensitive ($z\approx 600$) \cite{2012PhRvD..85d3522F}.
Different DM masses and annihilation channels yield different values of $f_\mathrm{eff}$ \cite{2009PhRvD..80d3526S}.
The \textit{Planck} analysis focuses on masses $\geq 5~\mathrm{GeV}$, for which thermal DM freezes out well before BBN.
At lower masses, the effects of DM mass during BBN should be taken into account for a consistent analysis.

\section{Impact of combined effects on CMB anisotropies} \label{subsec:jointeffects}

\begin{figure*}
\includegraphics[width=0.46\textwidth]{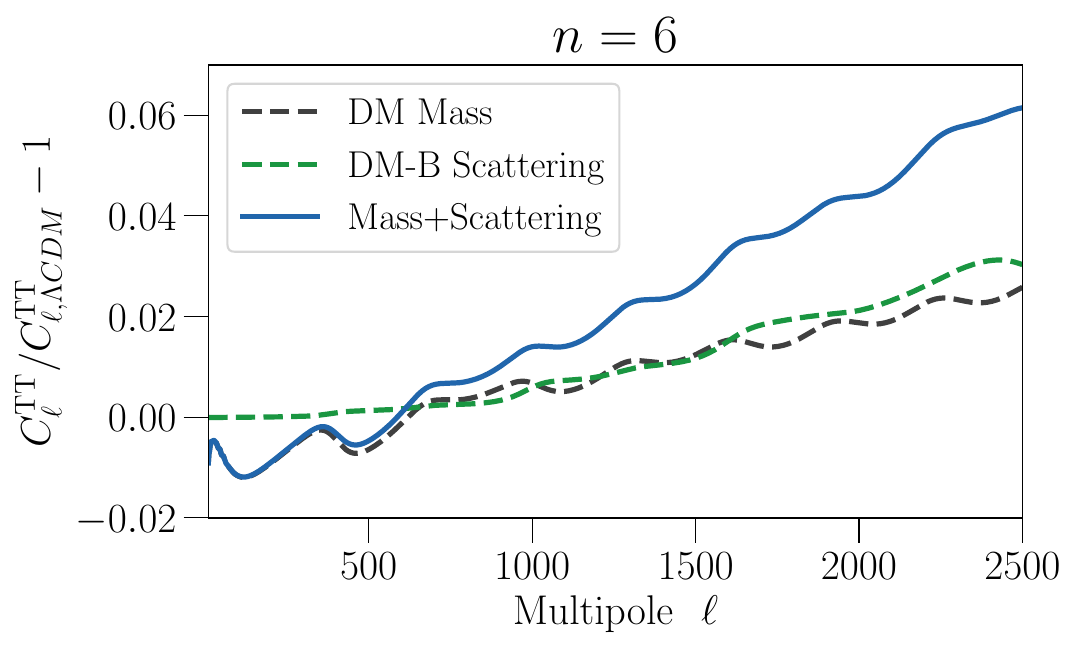}\ \ 
\includegraphics[width=0.46\textwidth]{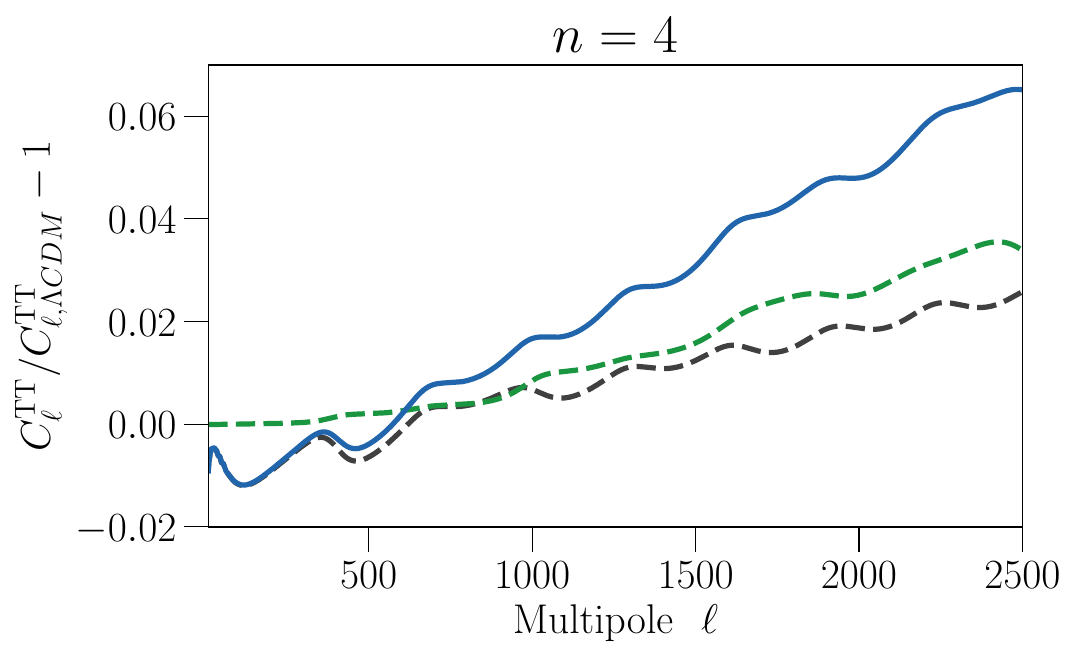}\\
\includegraphics[width=0.46\textwidth]{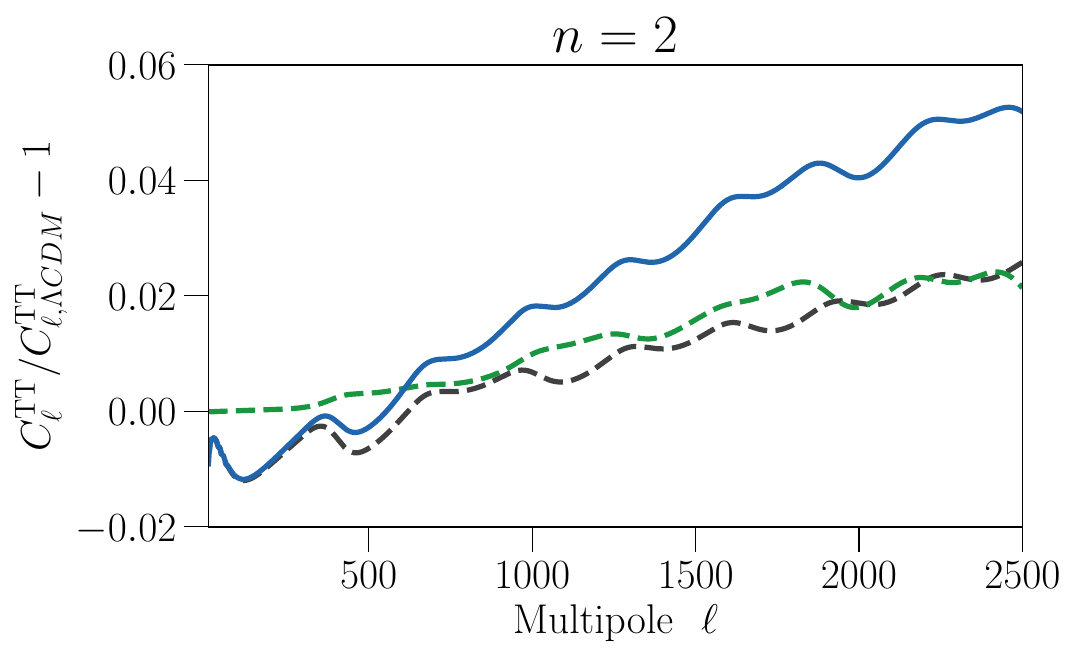}\ \ 
\includegraphics[width=0.46\textwidth]{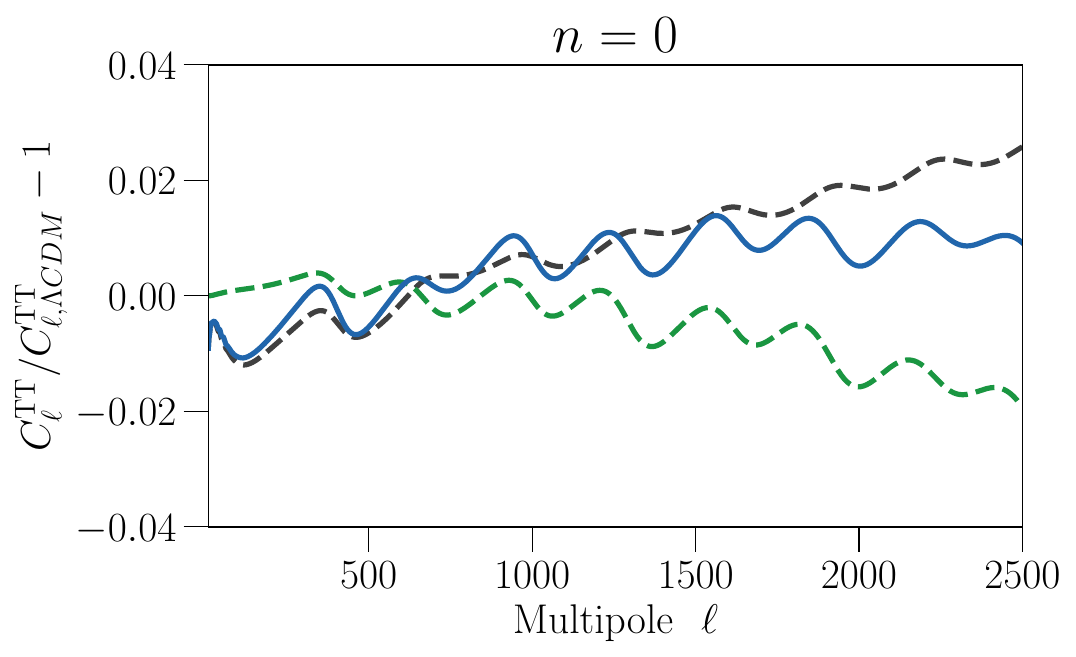}
\caption{Ratio of the CMB temperature power spectrum for different DM models, as compared to the CDM scenario, for $m_\chi = 10$ MeV. The gray dashed lines correspond to the effects of a light thermal-relic Majorana DM particle, electromagnetically coupled to SM particles in the early Universe. The green dashed lines capture the effects of DM-baryon elastic scattering models, with a momentum-transfer cross section that depends on the relative particle velocity, as a power law, with a power-law index $n$ denoted in the title of each panel. The cross section is fixed at the current 95\% CL CMB upper bound from Ref.~\cite{2021PhRvD.104j3521N}. The blue solid lines combine the effects of DM mass and DM-baryon elastic scattering. Note that all the other six standard cosmological parameters ($\Omega_bh^2$,$\Omega_\mathrm{dm}h^2$,$100\theta$,$\tau$,$n_s$,$A_s$) are fixed at their no-scattering best-fit \textit{Planck} values \cite{2020A&A...641A...6P}.}
\label{fig:cl}
\end{figure*}

In this section, we investigate the impact on CMB anisotropies due to the combined effects of a low DM mass and interactions DM has with SM particles.
We begin by considering two scenarios for the interactions: one in which DM elastically scatters with baryons and another in which DM has late-time residual annihilation.
The first scenario assumes there is no late-time annihilation, relevant for asymmetric DM models.
The second scenario assumes the effect of scattering is negligible.
We then turn to a specific model in which DM possesses an electric and magnetic dipole moment in order to study the simultaneous effects of DM scattering with baryons and late-time annihilation.

\subsection{DM mass and DM-baryon scattering} \label{subsec:M-IDM}

In order to compute CMB power spectra in a scenario that simultaneously accounts for the effects of light thermal-relic DM mass and elastic scattering with baryons, we implement the contributions of thermally coupled DM (through its effects on $Y_\mathrm{p}$ and $N_\mathrm{eff}$) into a modified Boltzmann code \texttt{CLASS} developed for a cosmology that features DM and baryon scattering \cite{2018PhRvD..98h3510B}.
In the original version of this code, the value of $Y_\mathrm{p}$ is usually treated using standard BBN, where $Y_\mathrm{p}$ depends only on baryon density $\Omega_b h^2$ and extra relativistic species $\Delta N_{\nu}$;
and $N_\mathrm{eff}$ is a constant parameter, with a default value of 3.044 \cite{2020JCAP...12..015F,2021JCAP...04..073B,2020JCAP...08..012A}.
In order to account for the effects of light thermal-relic DM mass $m_\chi$, we first use the publicly-available \texttt{AlterBBN} code \cite{2012CoPhC.183.1822A, 2018arXiv180611095A} to generate new BBN predictions for $Y_\mathrm{p}$, as a function of $\Omega_b h^2$, $\Delta N_{\nu}$, and $m_\chi$; and for $N_\mathrm{eff}$, as a function of $m_\chi$. Then we update the BBN interpolation table and related BBN calculation for $Y_\mathrm{p}$ within \texttt{CLASS} code, as well as the calculation of $N_\mathrm{eff}$.
In this work, we do not consider other relativistic degrees of freedom and set $\Delta N_{\nu} = 0$.

Fig.~\ref{fig:cl} illustrates the impact of DM mass, DM-baryon scattering, and their joint effects on the CMB power temperature spectrum; the power spectra shown in this Figure are normalized to the standard $\Lambda$CDM model.
In all the panels, we assume DM is a 10 MeV Majorana fermion that couples electromagnetically during BBN.
As studied in Ref.~\cite{2022JCAP...07..002A}, the presence of an electromagnetically coupled particle with mass between $2~\mathrm{MeV} \lesssim m_{\chi} \lesssim 20$ MeV can slow down the expansion rate, increasing the time available for neutron decay but also leading to a slower conversion rate between neutrons and protons; these two effects nearly cancel out, resulting in a slightly reduced helium abundance $Y_\mathrm{p}$.
Meanwhile, DM annihilating into photons after neutrino decoupling heats photons relative to the neutrinos, reducing the present-day ratio of neutrino-to-photon temperature, leading to a smaller value of $N_\mathrm{eff}$.
As a result, this 10 MeV particle eventually leads to an enhancement of the CMB power spectrum at smaller scales (see the gray dashed lines).

The effects of DM-baryon scattering are model-dependent \cite{2018PhRvD..97j3530X,2018PhRvD..98h3510B}, as shown in Fig.~\ref{fig:cl} (green dashed lines), where we fix DM mass $m_{\chi} = 10$ MeV and set the cross section $\sigma_0$ to its respective 95\% CL upper limits derived from a previous CMB analysis \cite{2021PhRvD.104j3521N}.
Scattering between DM and baryons leads to enhanced CMB power spectrum on the scales of interest in models with $n\geq 2$, while scattering with $n=0$ leads to a suppression of power at high multipoles.
We then combine the effects of DM mass and DM-baryon scattering (blue lines).
For $n\geq 2$, these two effects add up, leading to a larger deviation from the standard CDM scenario, which should result in more stringent bounds; for $n = 0$, these two effects cancel out in parts of the parameter space and should weaken the CMB bounds for a range of mass around 10 MeV (see Sec.~\ref{sec:mcmc-results}).

\subsection{DM mass and residual DM annihilation} \label{subsec:M-Anni}

\begin{figure}
\includegraphics[width=0.46\textwidth]{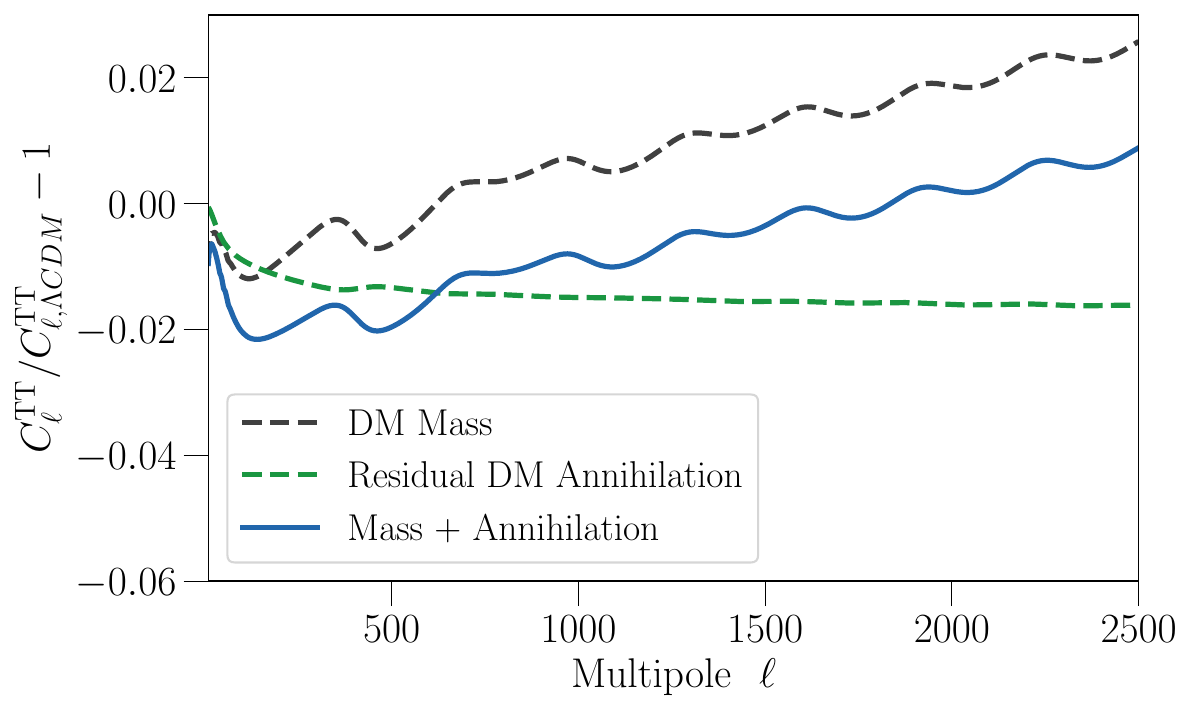}
\caption{Ratio of the CMB temperature power spectrum, compared to the CDM scenario, for $m_\chi = 10$ MeV. The gray dashed line corresponds to the effects of a light thermal-relic Majorana DM particle, electromagnetically coupled to SM particles in the early Universe. The green dashed line captures the effects of late-time residual DM annihilation, with the cross section fixed at the current 95\% CL CMB upper bound from Ref.~\cite{2020A&A...641A...6P}. The blue solid line combines the effects of a light DM mass and late-time DM annihilation. Note that all the other six standard cosmological parameters ($\Omega_bh^2$,$\Omega_\mathrm{dm}h^2$,$100\theta$,$\tau$,$n_s$,$A_s$) are fixed at their best-fit \textit{Planck} values in standard model \cite{2020A&A...641A...6P}.}
\label{fig:cl_anni}
\end{figure}

Using the same methods presented in previous section \ref{subsec:M-IDM}, we update the latest public version of \texttt{CLASS} \cite{2011arXiv1104.2932L,class2011} with added capability of simultaneously accounting for effects of light thermal-relic DM mass and late time residual DM annihilation.
The resulting CMB power spectra are shown in Fig.~\ref{fig:cl_anni}.

Residual DM annihilation reduces the CMB temperature power spectrum in a nearly scale-invariant manner at multipoles $\ell \gtrsim 500$ (green dashed line).
Thus, the enhanced power spectrum due to the effects of DM mass (gray dashed line) experiences an overall suppression when incorporating the effects of residual annihilation (blue solid line).

\subsection{Dipole DM}\label{sec:dipoleDM}

\begin{figure}
\includegraphics[width=0.44\textwidth]{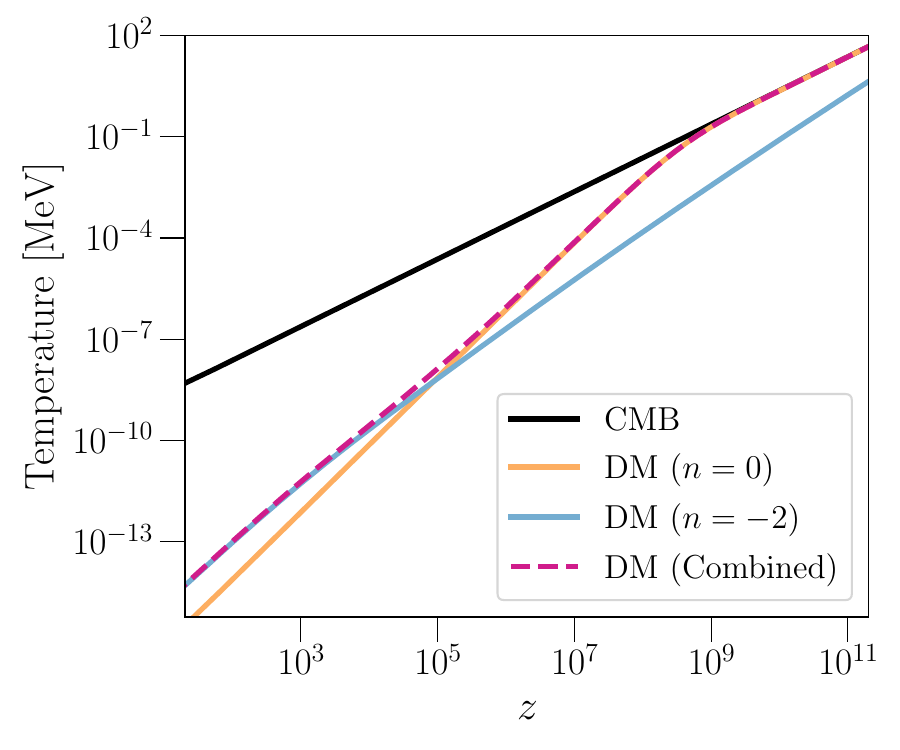}\\ 
\includegraphics[width=0.43\textwidth]{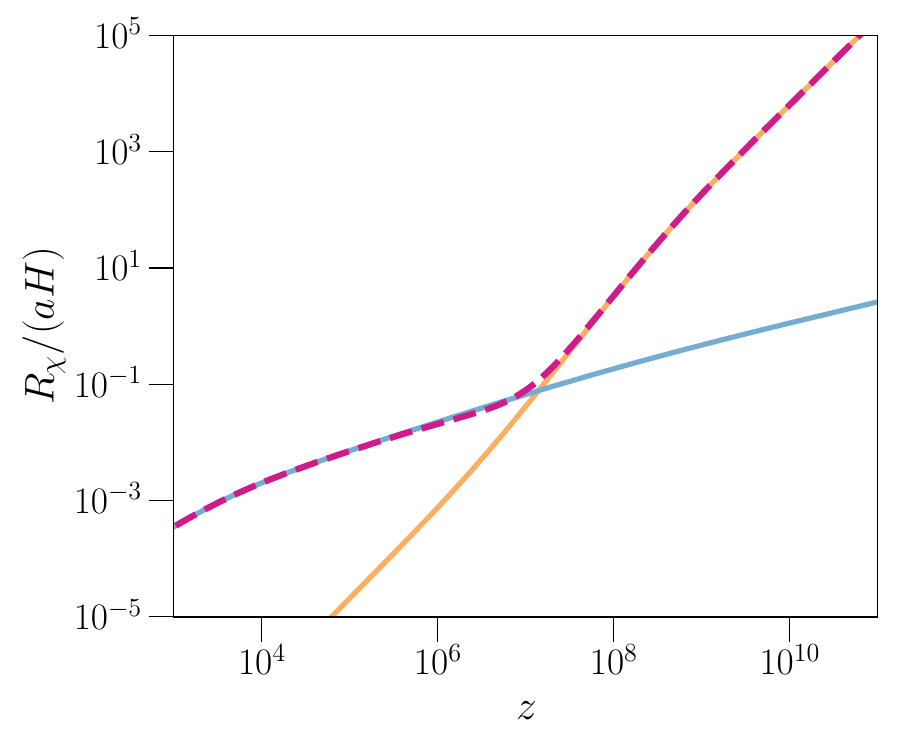}
\caption{Same as Fig.~\ref{fig:Tdm}, but for $n=0$ (orange line) and $n=-2$ (blue line) models; The dashed purple line represents the dipole DM model, which has a cross section containing $n=0$ and $n=-2$ terms. The cross sections are fixed at the 95\% CL upper bound from the \textit{Planck} analysis of the combined model with all three effects from DM mass, DM-B scattering and late-time residual DM annihilation, presented in this work.}
\label{fig:Tdm_combined}
\end{figure}

\begin{figure*}
\includegraphics[width=0.46\textwidth]{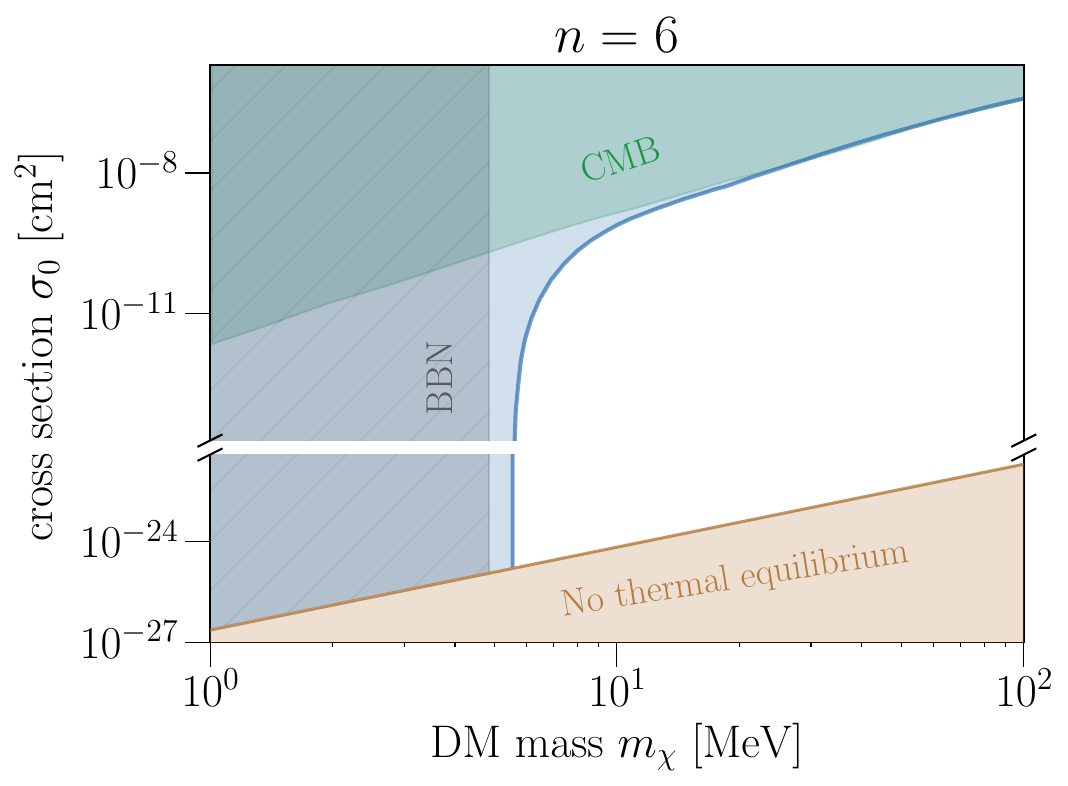} \ \ 
\includegraphics[width=0.46\textwidth]{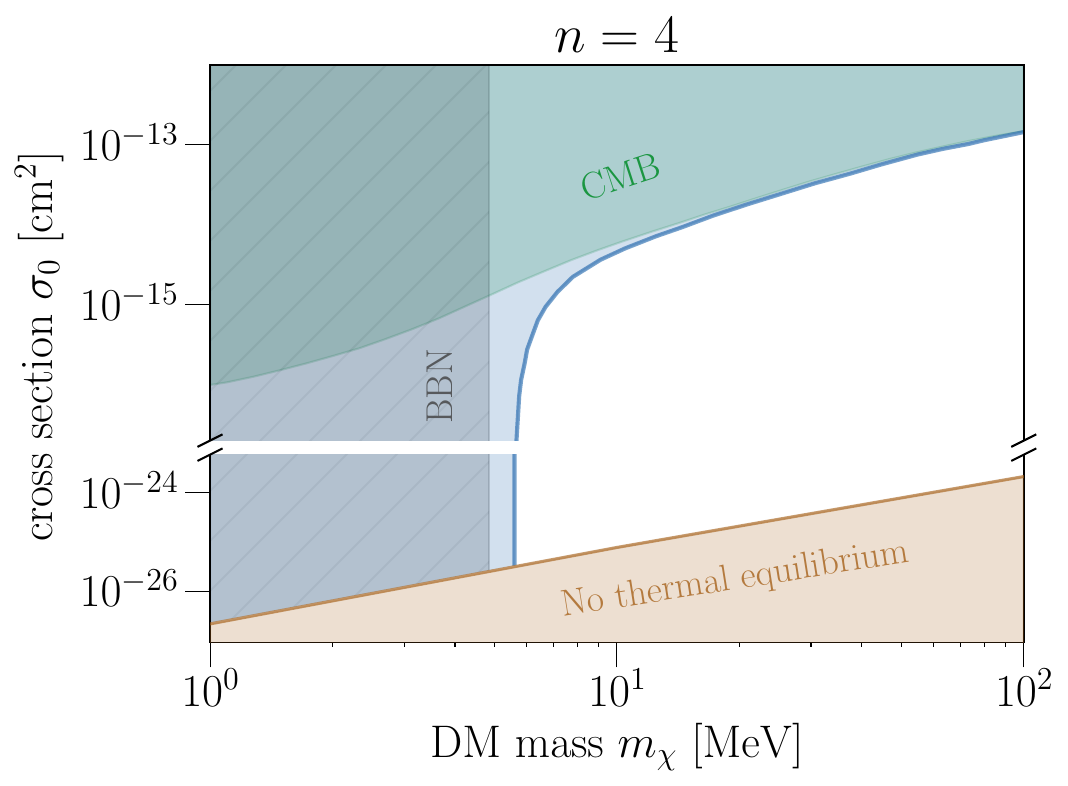}\\
\includegraphics[width=0.46\textwidth]{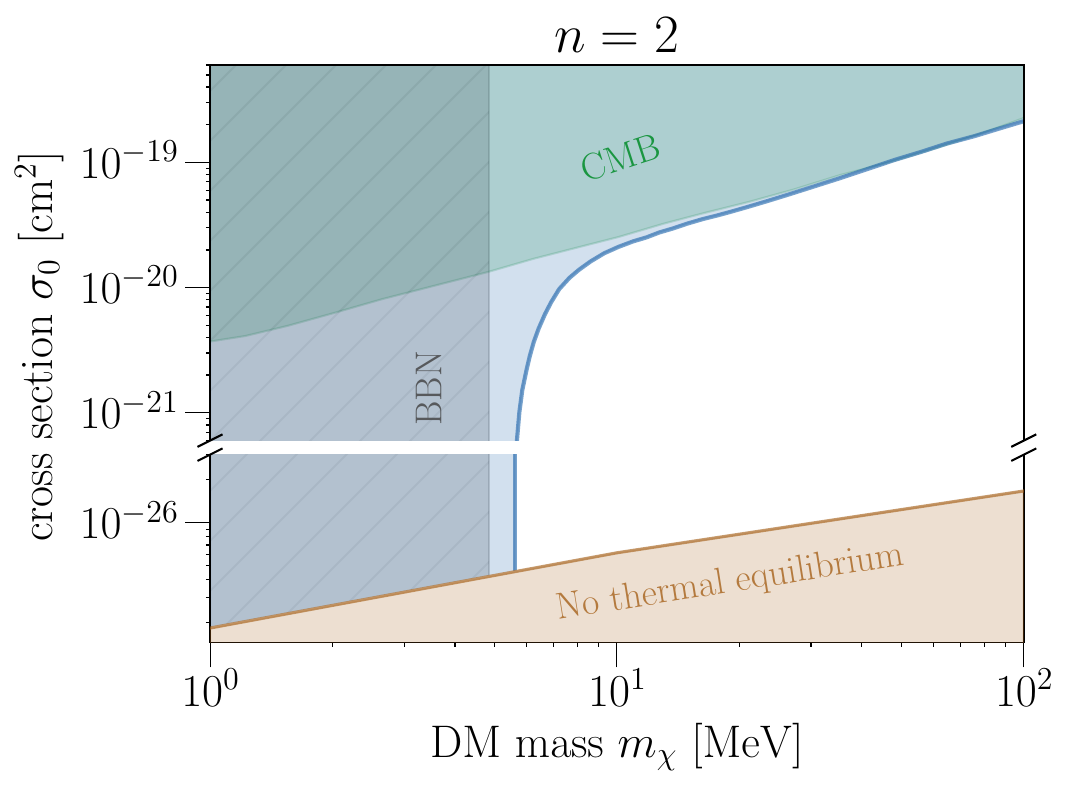} \ \ 
\includegraphics[width=0.46\textwidth]{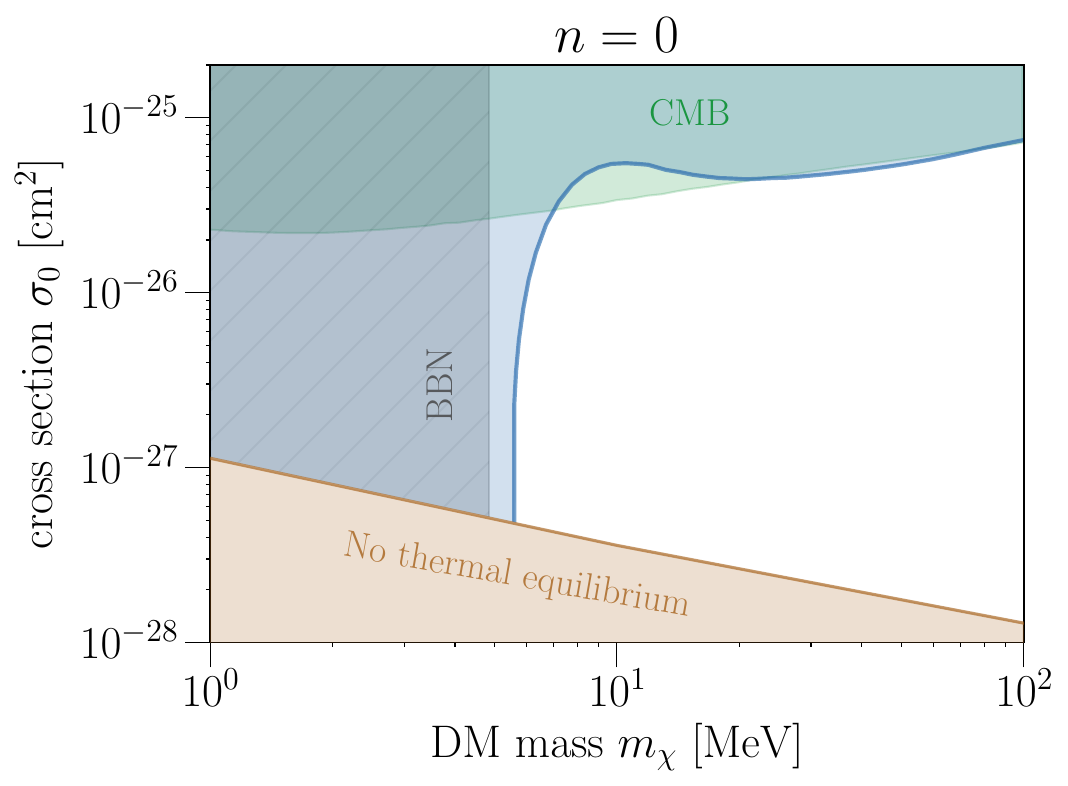}
\caption{Joint constraints on DM mass and elastic scattering cross section with protons. The blue shaded regions of the parameter space are excluded at 95\% CL by our analysis of \textit{Planck} 2018 data. We also present the CMB constraints on the cross section $\sigma_0$ in the cases that only consider the effects of DM-baryon scattering (green regions). The gray hatched regions correspond to the lower bounds on DM mass that arise from light thermal-relic particles that impact BBN from Ref.~\cite{2022JCAP...07..002A}.
We note that the joint consideration of the effects of mass and elastic scattering produces more stringent constraints than previous analyses for masses below $\sim20$ MeV (except for the constraints around 10 MeV in the case of $n = 0$). The lower brown regions represent the ranges of parameters in which thermal equilibrium with the SM during BBN is not maintained.}\label{fig:planck}
\end{figure*}

For simple scenarios in which a single operator governs the interaction between DM and the SM, the cross sections for DM annihilation and scattering are related by a crossing symmetry.
However, it is possible to have multiple interaction channels that allow the (late-time) annihilation and scattering cross sections to be treated independently, and we use the model of dipole DM as an illustrative example.

We consider DM with magnetic and electric dipole moments $\mathcal{M}$ and $\mathcal{D}$, respectively.
The effective Lagrangian describing interactions between DM and the SM is
\begin{equation}
  \mathcal{L}_\textrm{int} = -\frac{i}{2} \bar{\chi} \sigma_{\mu \nu}
  \left(\mathcal{M} + \gamma_{5} \mathcal{D}\right) \chi F^{\mu \nu} \ ,
\end{equation}
where $F^{\mu\nu}$ is the electromagnetic field strength and $\sigma_{\mu\nu} \equiv \frac{i}{2} \left[ \gamma_{\mu}, \gamma_{\nu} \right]$.
The scattering cross section with a charged fermion with mass $m_f$ is~\cite{2004PhRvD..70h3501S}
\begin{equation}
  \sigma_\textrm{MT} = \alpha \left[\frac{2\mathcal{D}^2}{v^2} + \mathcal{M}^2
    \left(3 - \frac{m_f (m_f + 4 m_{\chi})}{\left(m_f + m_{\chi}\right)^2}
    \right) \right] \ ,
\end{equation}
where $\alpha$ is the fine structure constant.
The electric dipole scattering cross section has a $v^{-2}$ dependence, while the magnetic scattering cross section is independent of relative velocity. We focus on a scenario where DM scattering with both proton and electron.
As shown in the upper panel of Fig.~\ref{fig:Tdm_combined}, the contribution from velocity-independent scattering allows DM to be in thermal equilibrium with the SM during BBN. 
The momentum-transfer rate $R_{\chi}$ (purple dashed line) shown in the lower panel of Fig.~\ref{fig:Tdm_combined} is simply the sum of the momentum-exchange rates that arise from the two different interactions.
Scattering with $n=0$ dominates at early times, while scattering with $n=-2$ dominates at late times.

Annihilation into SM particles occurs through both the electric and magnetic dipole interaction.
However, the electric dipole interaction leads to $p$-wave annihilation, which is subdominant to the magnetic dipole $s$-wave annihilation for $v\ll 1$.
The $s$-wave annihilation cross section is~\cite{Chu:2018qrm}
\begin{equation}
  \left\langle \sigma v \right\rangle = \frac{\alpha Z^2 \mathcal{M}^2}{2}
  \left[ 2 + \left(\frac{m_f}{m_\chi}\right)^2 \right]
  \sqrt{1 - \left(\frac{m_f}{m_\chi}\right)^2} \ .
\end{equation}
Since we are interested in low DM masses, we anticipate the relevant annihilation channel to be $e^+e^-$, such that $m_f = m_e$.

\section{Results}\label{sec:constraints}

In this section, we constrain the DM mass and its interactions using \textit{Planck} 2018 data.
We describe the data set and analysis method in section \ref{sec:data} and present numerical results in section \ref{sec:mcmc-results}.

\subsection{Data and method}\label{sec:data}

We use the most recent CMB aniostropy measurements from \textit{Planck} 2018 \cite{2020A&A...641A...6P}, including low-$\ell$ and high-$\ell$ multi-frequency temperature and polarization power spectra to place the joint bounds on DM mass and interactions.
We rely only on the lite (post-marginalized) likelihood for high-$\ell$ TTTEEE, since using the full likelihood does not lead to appreciable changes in the inferred parameter values for models we consider here. 
We perform series of Markov Chain Monte Carlo (MCMC) runs within the \texttt{Cobaya} sampling framework \cite{2021JCAP...05..057T, 2019ascl.soft10019T}.
We utilize \texttt{mcmc} sampler and employ the convergence criterion $R-1=0.01$, where R is the Gelman-Rubin threshold \cite{1992StaSc...7..457G}. 

In each MCMC run, we consider thermal-relic DM that couples electromagnetically to the SM during BBN.
For the DM-baryon scattering scenarios, we fix the power-law index $n$ of the DM-baryon scattering cross section and sample the posterior distributions of the six standard cosmological parameters (baryon density $\Omega_bh^2$, DM density $\Omega_\mathrm{dm}h^2$, acoustic scale $100\theta$, reionization optical depth $\tau$, scalar spectral index $n_s$, and amplitude of the scalar perturbations $A_s$), along with the DM mass $m_{\chi}$ and cross section coefficient $\sigma_0$.
For the case of late-time DM annihilation, we fix efficiency parameter $f_\mathrm{eff}=1$ and sample the posterior distributions of the six standard cosmological parameters, along with the DM mass $m_{\chi}$ and annihilation cross section $\left\langle \sigma v \right\rangle$.

\subsection{Numerical results}\label{sec:mcmc-results}

\begin{figure}
\includegraphics[width=0.45\textwidth]{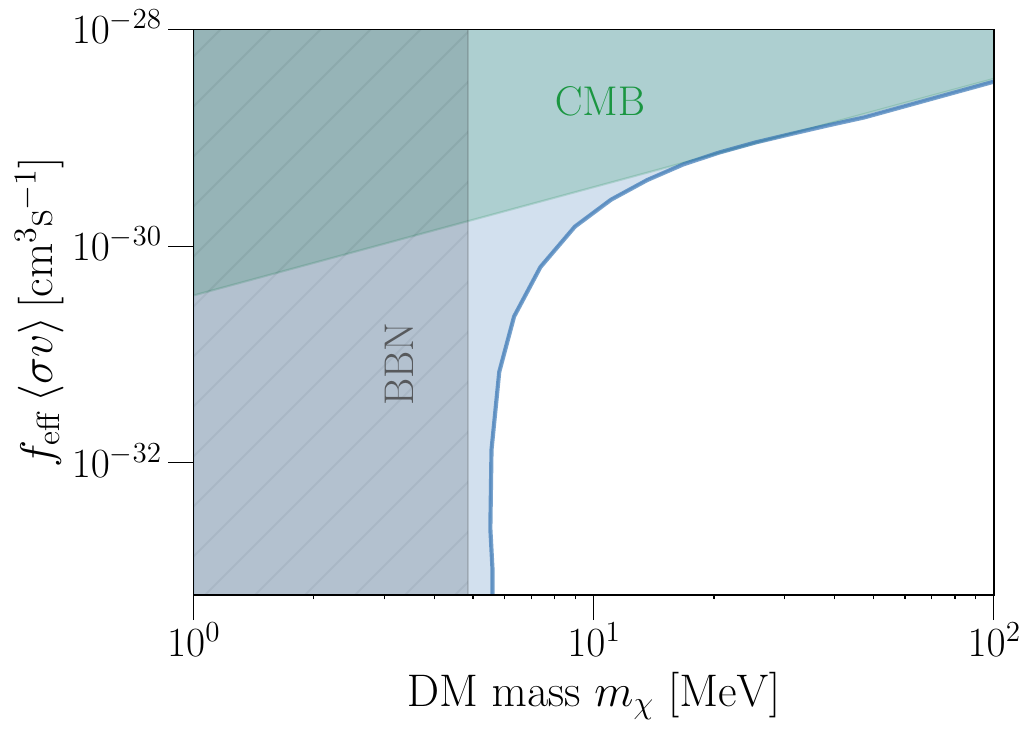}
\caption{Constraints on the annihilation cross section times the efficiency parameter $f_\mathrm{eff} \left\langle \sigma v \right\rangle$, and DM mass $m_{\chi}$. The blue region shows the parameter space excluded at 95\% CL by \textit{Planck} 2018 data. For comparison, we also show the constraints from previous work: the left hatched area corresponds to the lower bounds on DM mass that arise from light thermal-relic particles that impact BBN from Ref.~\cite{2022JCAP...07..002A}; and the upper blue area shows the parameter space excluded by the CMB measurements, giving $p_\mathrm{ann}<3.5\times 10^{-28} \mathrm{cm}^3\mathrm{s}^{-1}\mathrm{GeV}^{-1}$ at 95\% CL from Ref.~\cite{2020A&A...641A...6P}.
} 

\label{fig:anni}
\end{figure}

\begin{figure*}
\includegraphics[width=0.98\textwidth]{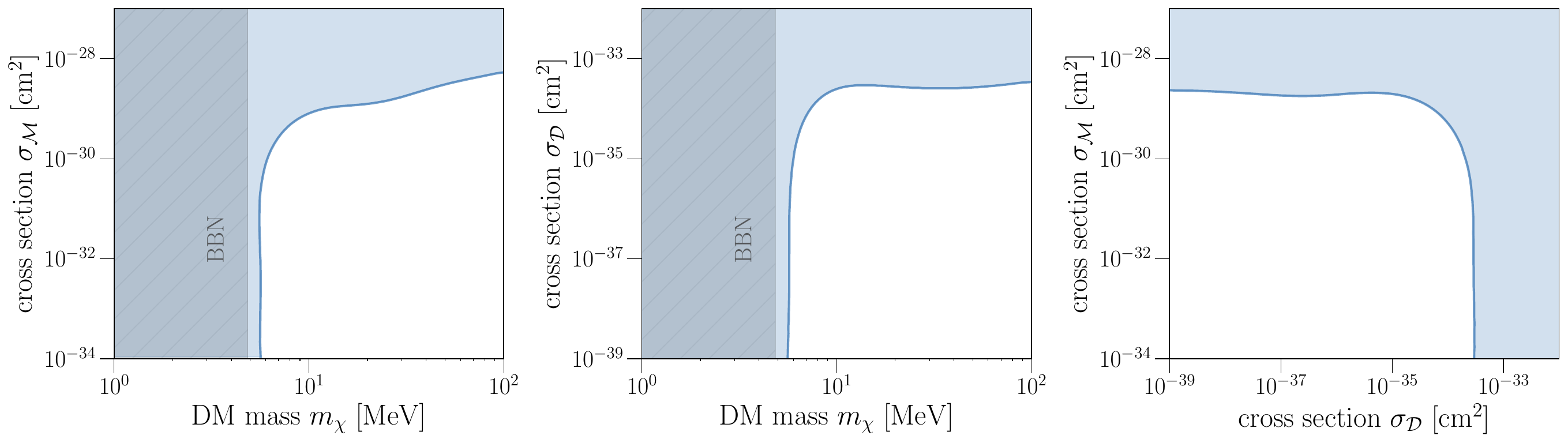}
\caption{Joint constraints on DM mass $m_{\chi}$ and cross sections $\sigma_\mathcal{M}$, $\sigma_\mathcal{D}$ in the combined scenario with all three effects from light DM mass, DM-baryon scattering, and late-time DM annihilation. The blue shaded regions of the parameter space are excluded at 95\% CL by our analysis of \textit{Planck} 2018 data. We also present the lower bounds on DM mass that arise from light thermal-relic particles that impact BBN from Ref.~\cite{2022JCAP...07..002A} (gray hatched area).}\label{fig:planck_combined}
\end{figure*}

The resulting constraints on $m_{\chi}$ and $\sigma_0$ for the models simultaneously accounting for the effects of light thermal-relic DM mass and DM-baryon scattering with $n\geq 0$ are presented in Fig.~\ref{fig:planck}.
The blue regions are excluded with 95\% CL by \textit{Planck} analysis. In the same Figure, we also present the CMB constraints on $\sigma_0$ when DM-baryon scattering effects are considered without consideration of the thermal-relic particle mass (green regions). The gray hatched regions correspond to the BBN bounds on DM mass from Ref.~\cite{2022JCAP...07..002A}.
For DM with $m_\chi \gtrsim 20$~MeV, there is no significant effect of the DM mass on CMB constraints, because freeze-out is complete prior to the decoupling of the SM neutrinos and therefore does not affect the process of BBN.
We find that in the range of thermal–relic masses relevant for BBN (at $m_{\chi} \lesssim$ 20 MeV), the DM mass and its scattering cross section are degenerate to each other (see Fig.~\ref{fig:cl}). The joint bounds are tighter than the ones from the individual analyses for all cases except one: for a DM mass of around 10 MeV and $n=0$, the effects of mass and scattering have some cancellation, as discussed in Section \ref{subsec:M-IDM}, thereby weakening the joint bound. 
Generally, modeling the effects of $m_{\chi}$ and $\sigma_0$ simultaneously can lead to an improvement in the sensitivity of CMB to the properties of light DM. Note that the constraints on $\Omega_bh^2$ in our combined models are tighter than the ones from Ref.~\cite{2022JCAP...07..002A}, which leads to more stringent constraints on $m_{\chi}$ even for very small cross sections, due to the positive degeneracy between $\Omega_bh^2$ and $m_{\chi}$, as discussed below. On the other hand, in order for the BBN mass bounds to hold, it is necessary that the DM stays in thermal equilibrium with the SM during BBN. The lower brown regions in Fig.~\ref{fig:planck} denote the parameter spaces for which DM interacts too weakly to guarantee thermal equilibrium during BBN; in that regime, the BBN bound is no longer valid.

The full posterior distributions for the six standard cosmological parameters and DM mass are presented in Fig.~\ref{fig:tri_DM-B} in Appendix \ref{appendix}.
Most of the parameters are strongly correlated with DM masses $5\ \mathrm{MeV} \lesssim $ $m_{\chi}\lesssim 20$ MeV, the mass regime that impacts the values of $Y_\mathrm{p}$ and $N_\mathrm{eff}$.
For example, there is a positive degeneracy between $\Omega_bh^2$ and $m_{\chi}$: DM annihilation to photons leads to a smaller value of $N_\mathrm{eff}$, while the value of $Y_\mathrm{p}$ is reduced in the mass range of 2--20 MeV \cite{2022JCAP...07..002A}---increasing $\Omega_bh^2$ can counteract these reductions \cite{2020A&A...641A...6P}.
This degeneracy is consistent with the inferred result in Ref.~\cite{2022JCAP...07..002A}, where they converted the CMB constraints on $Y_\mathrm{p}$ and $N_\mathrm{eff}$ to the constraints on $\Omega_bh^2$ and $m_{\chi}$.

We further perform MCMC analysis of CMB data for the scenario that simultaneously accounts for the effects of light thermal-relic DM and late-time DM annihilation.
Fig.~\ref{fig:anni} illustrates the resulting constraints on the annihilation cross section $\left\langle \sigma v \right\rangle$ times the efficiency parameter $f_\mathrm{eff}$, and DM mass. The blue area indicates the parameter space excluded at 95\% CL by \textit{Planck} data. The green region is ruled out by a previous CMB analysis for DM annihilation alone, giving $p_{ann} < 3.5 \times 10^{-28} [\mathrm{cm}^3\mathrm{s^{-1}}\mathrm{GeV}^{-1}]$ at 95\% CL \cite{2020A&A...641A...6P}.
The gray hatched region is ruled out by BBN alone~\cite{2022JCAP...07..002A}.
The effects on CMB anisotropies from DM mass and from residual DM annihilation are degenerate for $5\ \mathrm{MeV} \lesssim $ $m_{\chi}\lesssim 20$ MeV, leading to strong correlations between $m_{\chi}$ and $\left\langle \sigma v \right\rangle$. 
Modeling them simultaneously provides an improved constraint on the parameter space. The tighter bound on $m_{\chi}$ is caused by tighter constraint on $\Omega_bh^2$ as compared to the one from Ref.~\cite{2022JCAP...07..002A}. The posterior distributions for all the relevant parameters are presented in Fig.~\ref{fig:tri_anni} in Appendix \ref{appendix}, where we can clearly see a degeneracy between the cosmological parameters and the DM mass for $5\ \mathrm{MeV} \lesssim $ $m_{\chi} \lesssim 20$ MeV. 

The CMB analysis for the model that includes all three effects (light DM mass, DM-baryon scattering and late-time annihilation) is presented in Fig.~\ref{fig:planck_combined}. It illustrates the resulting constraints on the parameters of interest $\{m_{\chi}, \sigma_\mathcal{M}, \sigma_\mathcal{D}\}$, where $\sigma_\mathcal{M} = 2\alpha\mathcal{M}^2$ and $\sigma_\mathcal{D} = 2\alpha\mathcal{D}^2$ are the coefficients of the cross sections that correspond to DM-baryon scattering with $n=0$ and $n=-2$, respectively. The blue regions are excluded with 95\% CL by \textit{Planck} analysis, and the gray hatched areas correspond to the lower bounds on DM mass that arises from light thermal-relic particles that impact BBN. We can see that DM mass is degenerate with both cross sections ($\sigma_\mathcal{M}$ and $\sigma_\mathcal{D}$) in the range of BBN-relevant masses, and these two cross sections are also degenerated with each other. Considering the joint effects can improve the sensitivity of CMB measurements to individual parameters. The posterior distributions for all the other cosmological parameters are presented in Fig.~\ref{fig:tri_all} in Appendix \ref{appendix}.

\section{Summary}\label{sec:summary}

The presence of MeV-scale DM particles in thermal equilibrium with the SM plasma in the very early Universe has a number of cosmological consequences. The primary effect of a light thermal-relic DM on the CMB anisotropies is to change power at small scales, as a result of the impact of mass on $Y_\mathrm{p}$ and $N_\mathrm{eff}$ during BBN, while the effects of late-time interactions (elastic scattering with baryons and annihilation) can also alter matter disctribution on small scales.
In this work, we combine the early-time and the late-time effects of DM mass and interactions on CMB anisotropies and BBN.

We explore DM-baryon scattering with power-law dependence of the scattering cross section on the relative particle velocity, taking into account four values of the power-law index $n \in \{0, 2, 4, 6 \}$. We use a modified \texttt{CLASS} code with added capability of simultaneously accounting for effects of light thermal-relic DM mass and the interaction between DM and baryons.
The effects of the mass and interaction can add or even cancel out with each other, depending on the values of $n$.
We find that, in the range of thermal-relic masses relevant for BBN, the DM mass and cross section are degenerate with each other. Modeling them simultaneously provides more stringent bounds in almost all cases, leading to an improvement in the sensitivity of CMB to the properties of light DM. 
We then consider a scenario of light thermal-relic DM that features late-time annihilation.
We perform the CMB analysis of this joint model and provide improved constraints in the parameter space of DM mass and annihilation cross section.  Key results are shown in Figs.~\ref{fig:planck} and \ref{fig:anni}.

Finally, we present the first fully consistent bound on a simple DM model in which DM possesses an electric and magnetic dipole moment, in Fig.~\ref{fig:planck_combined}.
We simultaneously account for all three effects of light thermal-relic DM mass, DM-baryon scattering with $n=0, -2$, and residual DM annihilation. We find that the effects of mass are degenerate with the effects of scattering and annihilation; a joint analysis thus improves the sensitivity of CMB measurements to individual parameters. 

We expect that the analyses of the CMB measurements with more precision and accuracy at small scales, such as Simons Observatory \cite{2019JCAP...02..056A} and CMB-S4 \cite{2016arXiv161002743A, 2019arXiv190704473A}, are likely to yield even more stringent constraints on these DM models.
Furthermore, including other measurements from structure formation could help break degeneracies between the DM mass and interactions.
The same methods we employ in this study may be applied to other thermal-relic DM models probed by CMB anisotropies and other tracers of structure.
While for non-thermal relic models, e.g.\ millicharged DM, entirely new approaches are necessary to seek these candidates.
We leave such considerations for future work.

\section*{Acknowledgements}

VG and RA acknowledge the support from NASA through the Astrophysics Theory Program, Award Number 21-ATP21-0135.
VG acknowledges the support from the National Science Foundation (NSF) CAREER Grant No. PHY-2239205, and from the Research Corporation for Science Advancement under the Cottrell Scholar Program.
KB acknowledges support from the NSF under Grant No. PHY-2112884.

KB and VG gratefully acknowledge support from the Simons Center for Geometry and Physics at Stony Brook University, where part of the research for this paper was performed during the ``Lighting New Lampposts for Dark Matter and Beyond the Standard Model'' workshop.

\clearpage

\appendix
\onecolumngrid
\section{Full posterior distributions}\label{appendix}

\begin{figure*}[!h]
\includegraphics[width=0.98\textwidth]{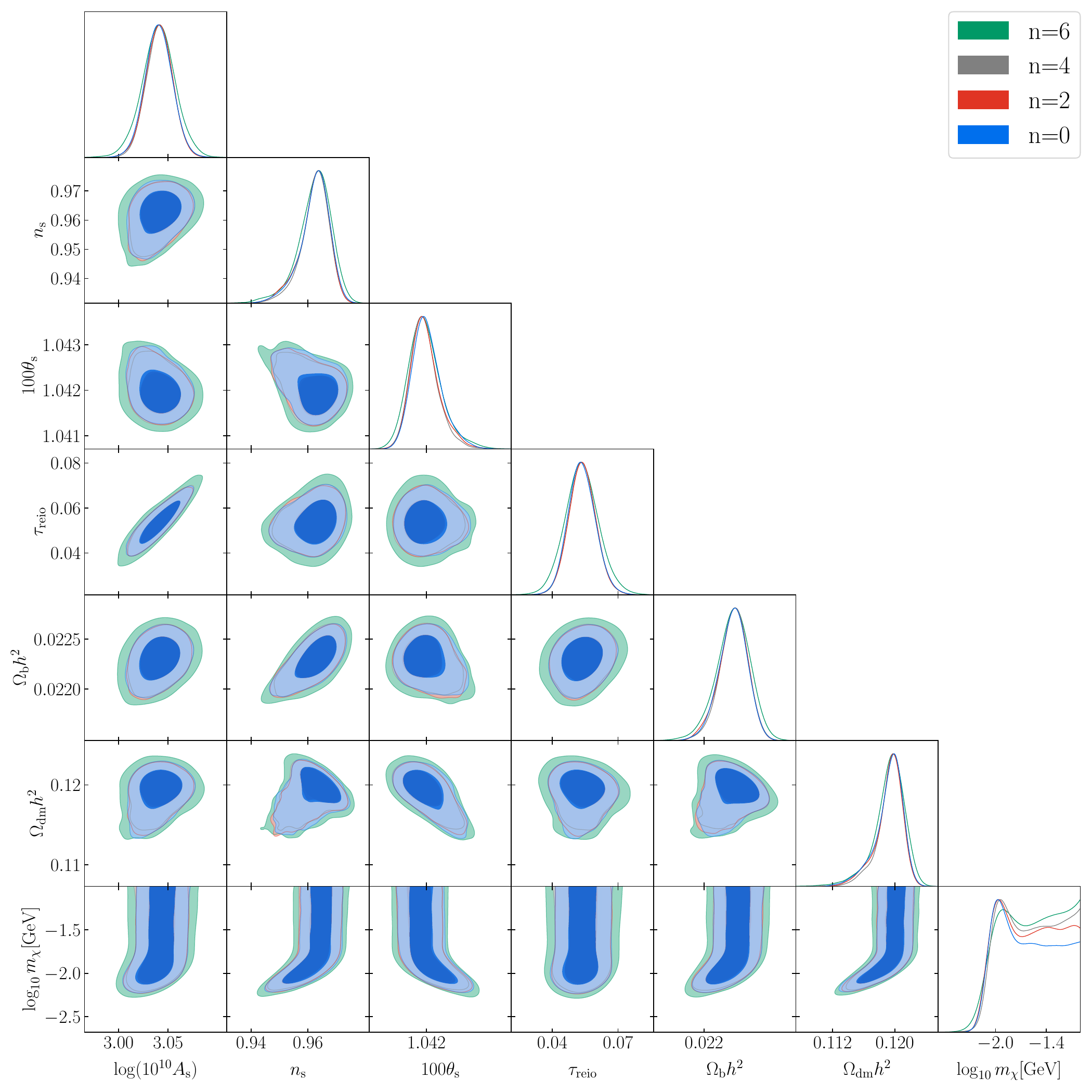}
\caption{The posterior probability distribution for the standard cosmological parameters and DM mass in different DM-baryon scattering models. We show the 68\% and 95\% CL contours, obtained from a joint analysis of Planck 2018 temperature and polarization anisotropies. The one-dimensional marginalized posteriors are shown at the top of each column.}\label{fig:tri_DM-B}
\end{figure*}

\begin{figure*}
\includegraphics[width=0.98\textwidth]{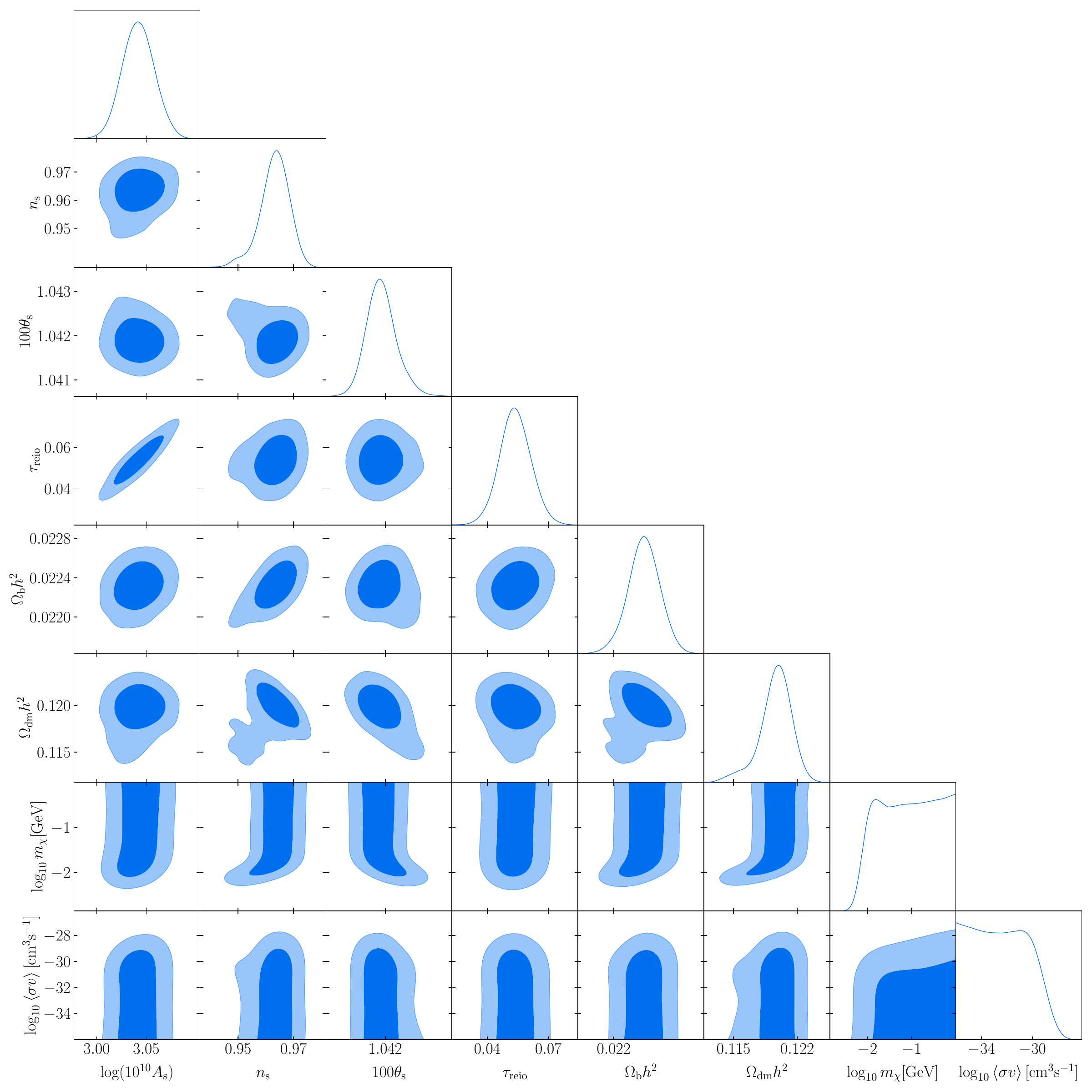}
\caption{The posterior probability distributions for the standard cosmological parameters, DM mass and the annihilation cross-section $\left\langle \sigma v \right\rangle$ , with the efficiency parameter $f_\mathrm{eff}$ set to 1. We show the 68\% and 95\% CL contours, obtained from a joint analysis of Planck 2018 temperature and polarization anisotropies. The one-dimensional, marginalized posteriors are shown at the top of each column.}\label{fig:tri_anni}
\end{figure*}

\begin{figure*}
\includegraphics[width=1\textwidth]{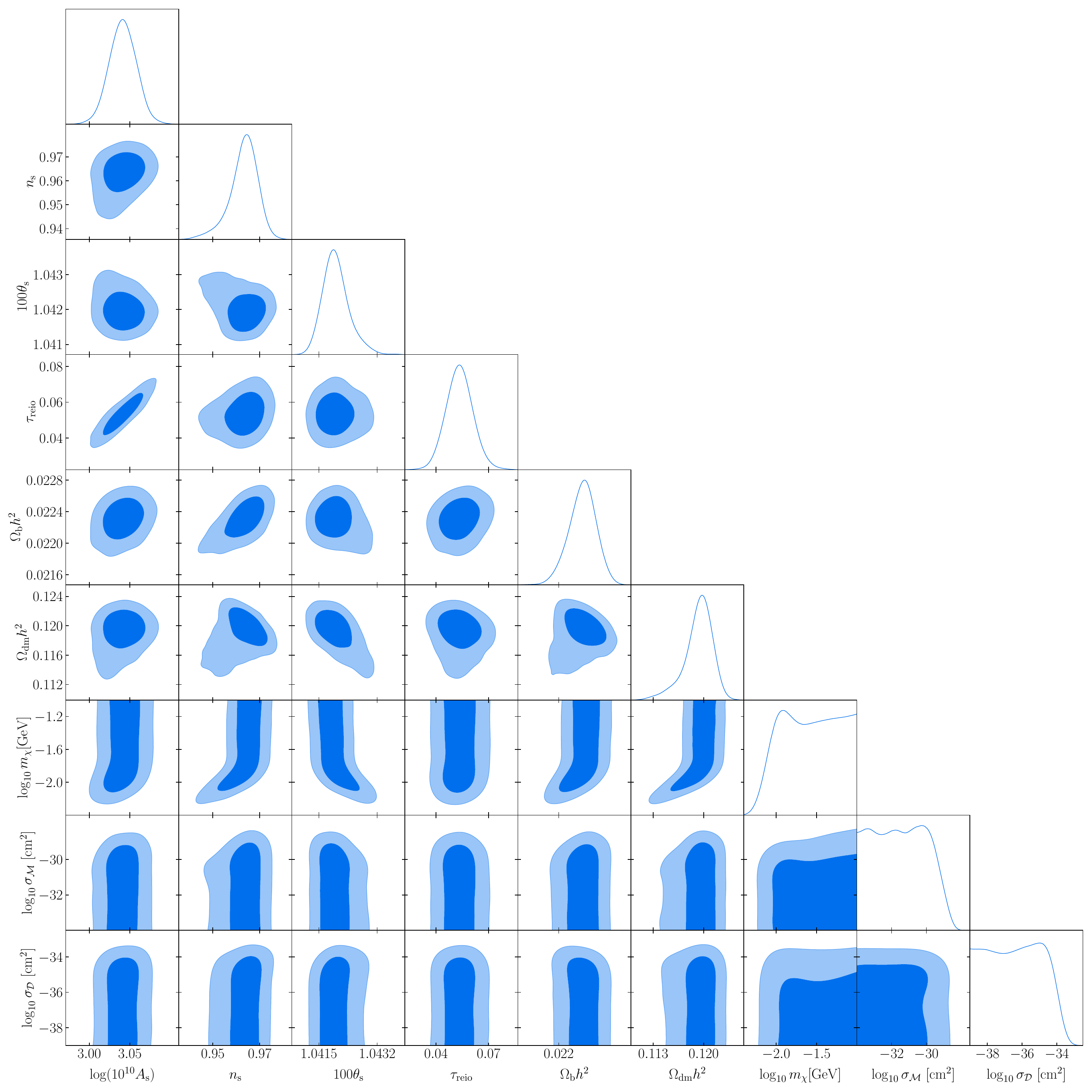}
\caption{The posterior probability distributions for the standard cosmological parameters, DM mass and cross sections $\sigma_\mathcal{M}$, $\sigma_\mathcal{D}$, in the combined scenario with all three effects. We show the 68\% and 95\% CL contours, obtained from a joint analysis of Planck 2018 temperature and polarization anisotropies. The one-dimensional, marginalized posteriors are shown at the top of each column.}\label{fig:tri_all}
\end{figure*}

We show the full marginalized posterior distributions for the relevant parameters in three scenarios i) with the effects of light DM mass and DM-baryon scattering,  in Fig.~\ref{fig:tri_DM-B}; ii) with the effects of light DM mass and late-time DM annihilation, in Fig.~\ref{fig:tri_anni}; iii) with all three effects of light DM mass, DM-baryon scattering and late-time DM annihilation, in Fig.~\ref{fig:tri_all}. 

\clearpage
\twocolumngrid

\bibliographystyle{IEEEtran}
\bibliography{refs}

\end{document}